\documentclass{emulateapj}
\usepackage{graphicx}
\usepackage{natbib}

\shorttitle{ }
\shortauthors{ }

\begin{document}
\title{The Origin of Color Gradients in Early-Type Systems and Their Compactness at High-z}

\author{La Barbera, F. \altaffilmark{1},  
        	de Carvalho, R.R.\altaffilmark{2}
}
\affil{INAF -- Osservatorio Astronomico di Capodimonte, Napoli, Italy, }
\affil{Instituto Nacional de Pesquisas Espaciais - INPE/MCT, Sao Paulo, Brazil}

\begin{abstract}
In this  ${Letter}$, we  present mean optical+NIR color  gradient estimates
for $5080$  early-type galaxies  (ETGs) in  the $grizYJHK$ wavebands
of the Sloan Digital Sky  Survey (SDSS) plus UKIRT Infrared Deep Sky
Survey (UKIDSS).  The color gradient is estimated as the logarithmic slope 
of the radial color profile in ETGs. With such a large sample size, we study the variation of the mean color gradient as a  function  of waveband with unprecedented  accuracy.   
We find that (i) color gradients  are mainly due, on average, to a metallicity 
variation of  about $ -0.4$dex per decade in galaxy radius;  and (ii)  a  small, 
but  significant, positive age gradient is present, on average, in ETGs, with
the  inner stellar population  being slightly  younger, by $\sim 0.1$dex per 
radial decade, than  the outer one.  Also,  we show  that the
presence of a  positive mean age gradient in ETGs, as  found in the present
study,  implies their  effective radius  to  be smaller  at high  $z$,
consistent with observations. 
\end{abstract}
\keywords{galaxies: clusters: general---galaxies: evolution---galaxi
es: fundamental parameters}


%

\section{Introduction}


Since the early 1970's, observations have shown that the color of ETGs 
reddens from the outskirts to the galaxy center (Faber 1972).  
Understanding the origin of such color gradient can strongly constrain the scenario 
of galaxy formation and evolution (see~\citealt{LMB04}).
Most studies  to date have been  plagued by  the large uncertainties due 
to small sample  sizes and short wavelength baselines where color gradients are derived. 
Despite these limitations they  suggest that color  
gradients in ETGs originate  from metallicity  variations  (Peletier et al.  1990; 
Tamura  et al.  2000; Tamura  \& Ohta  2003). Small, 
positive age gradients are also consistent with observations (e.g.~\citealt{SMG00}),
though their presence  has never been revealed in the family of ETGs as a whole.  
Positive age gradients are a robust prediction of the hierarchical paradigm of galaxy formation, as a consequence of the centrally peaked star formation of dissipative merging events.


Recent studies  have shown a  significant, intrinsic evolution  of
ETG sizes, with  high redshift galaxies being  more compact
 (by $\sim 50 \%$ at $z \sim 1$) than those of analogous stellar mass at $z \sim 0$. The
origin of this  variation in size is still a matter  of debate and may
be  the  result of  different  mechanisms (see  e.g.~\citealt{TRU09}),
specifically (i)  the effect of minor dry  mergers;  or (ii)  a puffing-up mechanism 
due to AGN  feedback~\citep{Fan08}; or (iii) the increased amount 
of dissipation involved in gas-rich mergers forming ETGs at high-redshift~\citep{KhS06}.

In  this Letter, we use  data  from the SDSS  and UKIDSS surveys to address the origin of the mean color  gradient in  ETGs in  the  nearby Universe  (z$<$0.1), connecting it  to the   observed  compactness   of   their  high-redshift counterparts~\footnote{We  assume   a  $\Lambda   \rm  CDM$   cosmology  with $\Omega_{\rm m}$ = 0.3,  $\Omega_{\Lambda}$ = 0.7, and $\rm H_{\circ}$
= 75 km $\rm s^{-1}$ $\rm Mpc^{-1}$.}. The  combination of  sample size,  homogeneity of data, and large wavelength  baseline, from optical to near-infrared (g through K bands), results  in an unprecedented  accuracy   in detecting both metallicity  and  age variations inside galaxies. In fact, while optical (SDSS) data are very sensitive to the effects of both metallicity (through line blanketing) and age, the NIR (UKIDSS) wavebands are dominated by the old, quiescent stellar populations, allowing the effects of age and metallicity to be effectively separated (see e.g.~\citealt{PVJ90}).

\section{The sample}
\label{polfit}

The  sample of  ETGs  was  defined  from SDSS-DR6  with
photometry and  spectroscopy available.   We selected all  galaxies in
the  redshift   range  of  0.05   to  0.095,  with   r-band  Petrosian
magnitude~\footnote{
      k-corrected       with      $
  kcorrectv4\_1\_4$~\citep{BL03}     through     rest-frame    filters
  blue-shifted  by  a  factor  $(1+z_0)$,
  adopting $z_0=0.1$ (see e.g.~\citealt{Hogg04}).}  M$_{r}{<}-20$.  The
choice    of    the   lower    redshift    minimizes   the    aperture
bias~\citep{GOMEZ03}, while the upper redshift limit makes the sample approximately  volume-complete, since the   value   of M$_{r}{=}-20$ roughly  corresponds to the apparent  magnitude limit of
the SDSS  spectroscopy ($r \sim  17.8$) at $z=0.1$.  Following  ~\citet{BER03a}, ETGs  are
those  galaxies  with  SDSS  parameters  $eclass  \!   <  \!   0$  and
$fracDev_r \! >  \!  0.8$. Also, we selected  only those galaxies with
spectroscopic warning  flags set to zero, and  with available velocity
dispersion, in  the range of  $70$ to $420$ km/s.   These requirements
yield a sample of $39,993$ ETGs. We matched this sample to the  
fourth  data   release  of the UKIDSS--Large Area Survey~\citep{Law07},  which provides NIR
photometry in the  $YJHK$  bands over a sky region significantly overlapping the SDSS. The
matching was done considering only frames with the better  quality flag ($ppErrBits <16$) in all bands.
For each ETG we selected the nearest UKIDSS detected galaxy within a matching radius~\footnote{The matching was done with the {\it CrossID} form of the WFCAM Science Archive (see http://surveys.roe.ac.uk/wsa/index.html for details). Changing the matching radius to $0.5$'' leads to decrease the sample size by only five galaxies, confirming the accuracy of the matching procedure.} of $1$''.
 $5080$ ETGs have  photometry available in all eight SDSS+UKIDSS
filters.

\section{Color gradient estimates}
\label{polfit}
The $grizYJHK$ images of each  galaxy were retrieved from the SDSS and
UKIDSS  archives, and processed  with 2DPHOT (see~\citealt{LBdC08}; hereafter LdC08).  
Galaxy images  were fitted  with
PSF convolved Sersic  models, allowing  us to  estimate structural
parameters,  i.e.  the  effective  radius,  $r_e$,  the  mean  surface
brightness  within that  radius, $< \! \mu \! >_e$,  and the  Sersic  index, $n$,
homogeneously  in  all  the   bands.  The PSF was accurately modeled~\footnote{For the PSF fitting in all wavebands, the mean value of the reduced $\chi^2$ distribution amounts to $\sim 1.04$, while the corresponding 90th percentile is $1.2$.} in all wavebands by fitting the five closest stars to each galaxy with a sum of three Moffat functions, 
as detailed in LdC08. Having  eight  wavebands,  seven
different color  gradients can be  estimated.  We used  the parameters
$r_e$,  $\mu_e$,  and  $n$,  to  estimate color  indices,  $g-X$  with
$X=rizYJHK$, as  a function of  the distance, $\rho$, from  the galaxy
center. For each waveband, we  estimate the mean surface brightness of
the de-convolved Sersic  model~\footnote{Using the de-convolved models makes the procedure insensitive to seeing effects. We also verified that color gradients do not correlate with PSF fitting $\chi^2$ values as well as with the error on the seeing FWHM, as estimated by 2DPHOT from the width of the {\it sure star locus} (see LdC08).} on a  set of concentric ellipses~\footnote{Ellipses are  equally  spaced  in equivalent  radius
$\rho$ by $0.01 r_{e,r}$ ,  where $r_{e,r}$ is the 
the  r-band $r_e$.},  whose ellipticity
and position angle  are fixed to the value  of the $r$-band
Sersic  fitting.     The color  index  $g-X$ at  a  given  radius $\rho$  is
obtained  by subtracting  the mean  surface brightness  values  at the
corresponding       ellipse.  Each color profile  is fitted in  the radial
range of  $\rho_{min} = 0.1  r_e$ to $\rho_{max}  = r_e$ (see e.g. ~\citealt{PVJ90}), by  using an
orthogonal least squares fitting procedure.  We estimate the  $g-X$ color 
gradients, $\nabla_{g-X}$, as the logarithmic   slopes   of   the  $g-X$   
profiles,   $\nabla_{g-X}={d  \!(g-X)}/{d \!(\log \rho)}$.

\begin{deluxetable}{c | c | c}
\tablecaption{Statistics of Color gradients.}
\tablehead{ $\nabla g-X$ & $\mu$ & $\sigma$ }
\startdata
g-r & -0.071 $\pm$ 0.003 & 0.121$\pm$ 0.004 \\
g-i & -0.084 $\pm$ 0.004 & 0137  $\pm$ 0.003 \\
g-z & -0.089  $\pm$ 0.006 & 0.173  $\pm$ 0.003 \\
g-Y & -0.231  $\pm$ 0.006 & 0.171  $\pm$ 0.004 \\
g-J & -0.279  $\pm$ 0.008 & 0.204  $\pm$ 0.005 \\
g-H & -0.284  $\pm$ 0.008 & 0.194  $\pm$ 0.004 \\
g-K & -0.297  $\pm$ 0.008 & 0.219  $\pm$ 0.005 
\enddata
\end{deluxetable}

\begin{figure}[t!]
\begin{center}
\includegraphics[height=80mm]{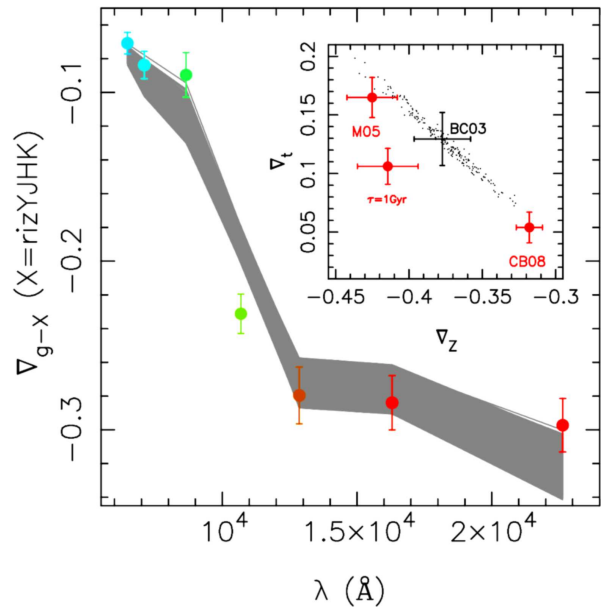}
\caption{ Mean  color gradients, $\nabla_{g-X}$ (filled circles),  as a function  of the
  effective wavelength of the filter $X$.   The  error bars mark  $2 \sigma$
  standard  errors.   The  grey  region  shows  the  set  of  age  and
  metallicity models that best match the observations (see text).  The
  dots in the inset plots the best-fitted  $\nabla_Z$  and $\nabla_t$ values obtained from 
   the CB03 SSP models by shifting mean color gradients according to their 
  uncertainties (see text). Variations in age and metallicity are generally correlated, 
  making the dots to align along a narrow direction in the plot. The error  bars are centered on the mean values of the best-fitted $\nabla_Z$  and $\nabla_t$ values,  and mark the corresponding $1 \sigma$ standard deviation.
   Different labels refer to the CB03 SSP model  (black), and to the M05, CB08, and  the $\tau=1$~Gyr  
  BC03 models  (red). 
\label{cgrad_lambda}
}
\end{center}
\end{figure} 

\section{Constraining age and metallicity variations}
\label{age_met_constrain}
Considering the effect of  age and metallicity, we write the equations:
\begin{eqnarray}
 \nabla_{g-X} & = & \frac{\partial (g-X)}{\partial {\log t}} \cdot \nabla_{t} + 
\frac{\partial (g-X)}{\partial {\log Z}} \cdot \nabla_{Z},
\label{grad_gX}
\end{eqnarray}
where  $\nabla_{Z}$  and   $\nabla_{t}$  are  the  logarithmic  radial
gradients of  age, $t$, and  metallicity, $Z$, across the  galaxy. The
quantities    $X_t={\partial   (g-X)}/{\partial    {\log    t}}$   and
$X_Z={\partial (g-X)}{\partial {\log  Z}}$ are the partial logarithmic
derivatives  of $g-X$  with  respect  to   age  and
metallicity. The above linear equations  hold if the color indices are
continuous  functions of  $t$  and  $Z$, and  the  absolute values  of
$\nabla_{Z}$  and  $\nabla_{t}$ are  small.  \\  From the  operational
viewpoint,  we  consider  two  stellar populations  and  describe  the
properties of  the galaxy at  the inner and outer  radii, $\rho_{min}$
and $\rho_{max}$, respectively.  Since our aim is that of characterizing stellar population gradients in the ETG population as a whole, we estimate the quantities
$\nabla_{Z}$ and $\nabla_{t}$ from Eq.~\ref{grad_gX}; using the mean values of the color
gradient  estimates~\footnote{We verified that applying the same procedure to the single color gradient estimates, and then computing the mean values of $\nabla_{Z}$ and $\nabla_{t}$  changes  these quantities by less than $0.02$dex. }  $\nabla_{g-X}$ (see Sec.~5).   We  accomplish  this  in  two
steps: (i) We parametrize the  colors available in terms of age and
metallicity  by  using different  spectrophotometric  codes.  We  take
simple   stellar   populations   (SSPs)   from~\citet{BrC03}   (BC03),
\citet{M05}  (M05), and  Charlot  and Bruzual  (2008, in  preparation;
CB08). These SSPs are based on different synthesis techniques and have
different  IMFs.  The  M05 model  uses the  fuel  consumption approach
instead of  the isochronal synthesis of  BC03 and CB08.  The CB08 code
implements a new AGB phase treatment~\citep{MG07}. The IMFs are: Scalo
(BC03), Chabrier (M05), and Salpeter  (CB08).  Moreover, we also use a
composite stellar  population model from BC03  having exponential star
formation rate  (SFR) with e-folding time of  $\tau=1$~Gyr~\footnote{Though 
the determination of absolute values of  age and metallicity at a given galaxy radius is 
beyond the scope of this paper, we have introduced the model with $\tau=1Gyr$ since it is 
able to match also the absolute values of central colors in ETGs~\citep{LBM03}.}. The models
are  folded with  the $grizYJHK$  throughput curves,  and  flux values
computed  for different  values  of $t$  and  $Z$. Here,  we use  ages
spanning  from   $7.8$  to  $12.6$Gyrs~\footnote{corresponding   to  a
  formation redshift of $z \sim 1.2$ and to the age of the Universe in
  the  adopted  cosmology},  and  metallicities  from  $0.2$  to  $2.5
Z_{\odot}$ for BC03 and CB08  models, and from $0.05$ to $2 Z_{\odot}$
for M05.  To calculate the color  derivatives with respect  to $t$ and
$Z$, for each waveband we  fitted the corresponding flux values with a
two-dimensional eight order  polynomial in $\log t$ and  $\log Z$. The
rms  of the  fits is  smaller  than $0.01$mag  for all  the bands  and
models. 
\\ (ii) We solve Eq.~\ref{grad_gX} in a $\chi^2$ sense,
by minimizing the expression
 \begin{eqnarray}
  \chi^2 & = & \sum_X \left( \nabla_{g-X} - X_t \nabla_t - X_Z \nabla_Z \right)^2. \label{chi2} 
\label{chieq}
 \end{eqnarray}
with respect to $\nabla_t$ and $\nabla_Z$. We evaluate the derivatives
$X_t$  and $X_Z$  at  a given  $t$  and $Z$  values, representing  the
average  age  and  metallicity  of  the two  stellar  populations.  
\section{Results}
\label{age_met_grad_sec}
Table 1 lists the statistics  of $g-X$ color gradients. We compute the
peak  value,  $\mu$, and  the  width,  $\sigma$,  using the  bi-weight
location estimator~\citep{Beers:90} taking $1000$ bootstrap iterations
to estimate  uncertainties. As  we can see,  the peak value  shifts to
more  negative values  as  we  move to  longer  wavelengths, with  a
minimum  $\mu  \sim  -0.3$  in  $g-K$, and the  width becomes progressively
larger  from the  $r$ to  the $K$ band. Because  we have many galaxies,  
the random errors  on $\mu$ and $\sigma$ are very small.

Fig.~\ref{cgrad_lambda} shows the mean color gradient, $\nabla_{g-X}$,
as  a function  of  the effective  wavelength  of the  filter $X$, where the trend
reflects how stellar population
properties  vary, on average,  as a  function of  radius in  ETGs.  We
adopt the procedure  described in Sec.~\ref{age_met_constrain}  to infer the
mean age  ($\nabla_t$) and metallicity  ($\nabla_Z$) radial gradients.
To this  end, we minimize  Eq.~\ref{chieq} for each of  the stellar
population   models  described  in   Sec.~\ref{age_met_constrain}.  We
estimate  the  quantities  $X_t$  and  $X_Z$  in  Eq.~\ref{chieq}  for
$t=10$~Gyr~\footnote{In   the  adopted   cosmology;  this   age  value
  corresponds to $z \sim  2.3$} and $Z=Z_{\odot}$. Then, inserting the
best-fit values of  $\nabla_t$ and $\nabla_Z$ into Eq.~\ref{grad_gX},
we derived  the color gradient values that  best-fit the observations.
We performed  $N=1000$ iterations,  shifting, each time,  the observed
mean color gradients according to the corresponding uncertainties. Each set
of   best-fitted  color   gradients  defines   a  polygonal   line  in
Fig.~\ref{cgrad_lambda},  obtained   by  connecting  the  best-fitting
values of $\nabla_{g-X}$. The grey region in the figure shows the area
occupied  by the  $N=1000$ bootstrap  solutions. The  inset  shows the
values  of  $\nabla_t$  and  $\nabla_Z$ obtained  from  the  different
iterations. Computing the average  and the standard deviation of these
values,  we obtain  $\nabla_t=0.13 \pm  0.02$ and  $\nabla_Z=-0.38 \pm
0.02$,   respectively.  
Using a different radial range for
computing color gradients, with $\rho_{min} = 0.05 r_e$ to $\rho_{max}
=  2  r_e$,  does  not  change  at all  the  values  of  the
metallicity  and  age  gradients.  We also  estimated  $\nabla_t$  and
$\nabla_Z$ by  computing $X_t$ and  $X_Z$ for different values  of $t$
and $Z$, varying $t$ between  $9$ and $12$~Gyrs, and $Z$ between $3/4$
and  $2 Z_{\odot}$. The  inferred absolute  value of  $\nabla_Z$ ranges
from $\sim 0.3$  to $ \sim 0.4$~dex, while the  value of $\nabla_t$ is
always positive,  varying between $\sim  0.01$ and $\sim  0.25$dex~\footnote{We
obtain the minimum value of $0.01$ only for the CB08 model, and only for the lowest
metallicity  case. For  the BC03  and M05  models, we  find $\nabla_t
\widetilde{>} 0.1$~dex regardless of the $t$  and $Z$ values. }. 

\section{Discussion}
\label{SIZE}
We studied mean color gradients in ETGs using an
unprecedentedly  large and homogeneous  sample with data from optical to 
NIR wavebands.  We find that, on
average, the main  driver of color gradients is  a radial variation of
metallicity, and  that a
small, positive, age gradient seems  to be present, on average, in ETGs, implying a
mildly younger stellar population to the galaxy center. The mean metallicity gradient varies between
$\nabla_Z = -0.32  \pm 0.02$ and $\nabla_Z=-0.425  \pm 0.02$,
depending  on the stellar-population  model adopted  to fit  the color
gradients.  Previous studies  based on much smaller samples,
found $\nabla_Z$ mostly between $-0.2$ and $-0.3$,
with a typical uncertainty  of $\sim 0.1$ (e.g.~\citealt{PVJ90, SMG00,
  IMP02,  LBM03,  TaO03, dPCD04,  Wu05}), consistent with what we find here. \\
Past literature has shown that age gradients do not explain color
gradients, although a small positive age gradient of $\nabla_t \sim
0.1$ is still consistent with observations (see e.g.~\citealt{SMG00,
  LBM03}). For instance, ~\citet{TaO05} find a value of $\nabla_t =
0.1 \pm 0.14$, while ~\citet{Wu05}, analyzing optical--NIR color
gradients for 36 nearby ETGs, report $\nabla = 0.02 \pm 0.04$. Here,
we are able to detect a small but significantly positive age gradient
in ETGs. The presence of younger stars in the center of ETGs is
expected in most hierarchical formation scenarios. During the merger of
gas-rich systems, gas dissipates its kinetic energy, falling into the
galaxy center and forming stars.\\ The detection of a positive age
gradient also adds new insight to the fact that ETGs
have smaller effective radii at higher redshift when compared to $z
\sim 0$ (e.g.~\citealt{DAD05, TRU06, LON07, CIM08, R08, vW08, vD08,
  B08, SAR09}). Since the luminosity evolution of a stellar population
is approximately independent of its metallicity, the internal
metallicity gradient of ETGs is assumed not to modify the effective
radius with redshift.  However, a difference in the formation epoch of
two stellar populations corresponds to larger and larger differences
in luminosity as we approach the formation epochs.
\begin{figure}[t!]
\begin{center}
\includegraphics[height=80mm]{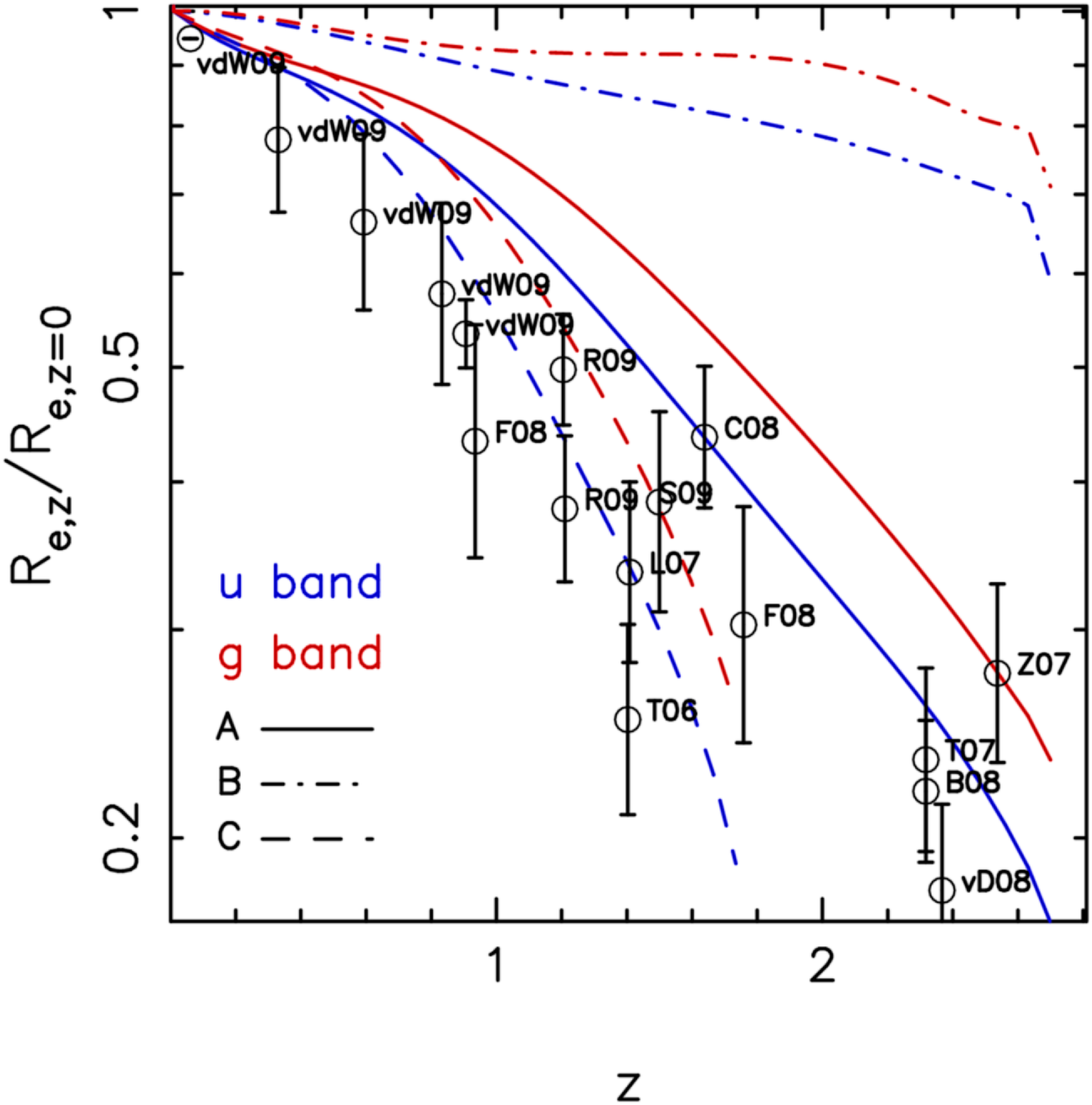}
\caption{ Evolution of  the effective radius of ETGs  as expected from
  different    radial gradients of age and metallicity.  Line  types  
  correspond to  different age  and metallicity  gradient models,  as shown  
  in the  lower--left  of the figure  (see the  text).  Circles with  
  1$\sigma$  error  bars show the observed evolution in size from previous 
  studies (\citet{vW08}, \citet{SAR09}). 
\label{size_ev}
}
\end{center}
\end{figure} 
Our result implies  that a younger stellar population  in the galaxy
center  would  brighten more  rapidly  with  redshift  than the  outer
stellar population, making the profile more concentrated 
at   high   $z$,   and   possibly  inverting   the   observed   colour
gradients.   To quantify  these effects, we take an $r^{1/4}$ law in the
same radial range of $\rho_{min}$ to $\rho_{max}$ where we derived the
internal  color gradients  (Sec.~\ref{polfit}).  At  the  inner radius
$\rho_{min}$,  we  assume  a  stellar  population  with  metallicity
$Z_{min}$  and formation redshift  $z_{f}$.  Then, for  a given  radius, the
profile is evolved with redshift according to the luminosity evolution
of  a  SSP  whose  age  and metallicity  are  computed  from  $z_{f}$,
$Z_{min}$,  $\nabla_Z$ and  $\nabla_t$. At a  given redshift, $z$, the profile is  fitted with an
$r^{1/4}$ law, yielding an effective  radius
$R_{e,z}$.  We used three models  with  different  values  of  $z_f$,
$\nabla_t$,  and  $\nabla_Z$.   We  fixed  $Z_{min}=Z_{\odot}$,  since
varying it in  the range of $1/2 Z_{\odot}$ to  $2 Z_{\odot}$ makes
$R_{e,z}$ change only a  few percent. All the evolved
profiles were well fit by the $r^{1/4}$ law,  with an rms smaller
than $\sim 0.05mag/arcsec^2$. Fig.~\ref{size_ev} compares the 
evolution in size, $R_{e,z}/R_{e,z=0}$, from different sources in the literature 
with that expected from the three  previously mentioned models. Since the data sample different restframes, from UV to optical wavebands, we constructed the models in both 
the $u$ and $g$ bands. Model A takes $z_f=2.8$ and
$\nabla_t=0.08$, an age gradient  consistent with our
findings and  such  that  the  stellar  population  at
$\rho_{max}$ has an age limited by that of the Universe. Model B uses the same
$z_f$  as  model A, but with $\nabla_t=0$ and $\nabla_Z=-0.3$. Model C has the same
metallicity  and age gradients as
Model A for BC03 SSPs,  but a younger inner stellar population
with $z_f=1.8$.  As  expected, a pure metallicity model  (Model B) does
not predict  any significant size  evolution with redshift. The evolution 
is mildly stronger in the restframe UV, with $R_{e,z}/R_{e,z=0}$  smaller in the 
u than in g band. At $z \sim 0$, the value of $\nabla_{u-g}$ is about 
$-0.2$~\citep{Wu05}. Following the method of~\citet{SpJ93}, this gradient implies that effective radii of ETGs increase by $\delta\sim 20 \%$ from $g$ to $u$ band at $z \sim 0$. The waveband 
dependence of $R_{e,z}/R_{e,z=0}$  would tend to reduce and possibly invert this trend,
producing $\nabla_{u-g}>0$ at high redshift. The inversion 
takes place at a redshift, $z_{I}$, where the ratio of $R_{e,z}/R_{e,z=0}$ between the g and u bands equals the value of $\delta$. For both models A and C, $z_{I}$
is very close to $z_f$, being $1.6$ and $2.4$, respectively. 
Interestingly, positive color gradients in high-redshift ETGs have been detected
by ~\citet{FER05}, who found that about one-third of field ETGs at $z \sim 0.7$ have 
blue cores, in contrast  to only  $10\%$  at  lower  redshift  (see
also~\citealt{MAE01, MEN04, FER09}). \\
Fig.~\ref{size_ev} shows that a  single  formation epoch for the inner stellar population of ETGs does not  reproduce the observations. Model A 
predicts a size  evolution of $\sim 80 \%$ at $z
\sim  2.4$, in  good  agreement with  the  data at $z>2$. However,  the
evolution is too  shallow for lower redshifts, where  a smaller $z_f$ is required  to fit  the data
(model C). This result could be explained in a picture whereby ETGs form 
at different redshifts by gas-rich mergers, and the redshift-size trend
is due to the increase in the amount of dissipation in
mergers with redshift~\citep{KhS06}. Fig.~\ref{size_ev} suggests that the central bursts of star formation take   place at different redshifts, above $z \sim 1.5$. However, the present 
analysis does not exclude a picture where the younger stars in the center form at very high 
redshift in most galaxies, with the color gradients accounting for only some part of the 
redshift-size  relation. 

\begin{acknowledgements}
We thank  G. Djorgovski, I. Ferreras, R. Gal, and P. Sarracco for several comments that helped to improve  this paper. We  thank the referee for helpful suggestions. We used data from the SDSS (http://www.sdss.org/collaboration/credits.html). This work is based on data obtained as part of the UKIRT Infrared Deep Sky Survey. 
\end{acknowledgements}

\end{document}